# A Novel Approach to Detect Phishing Attacks using Binary Visualisation and Machine Learning

Luke Barlow*, Gueltoum Bendiab†, Stavros Shiaeles†, Nick Savage†
*CSCAN, University of Plymouth, PL4 8AA, Plymouth, UK
luke.barlow@students.plymouth.ac.uk
†Cyber Security Research Group, University of Portsmouth, PO1 2UP, Portsmouth, UK
gueltoum.bendiab@port.ac.uk, sshiaeles@ieee.org, nick.savage@port.ac.uk

*Abstract*—Protecting and preventing sensitive data from being used inappropriately has become a challenging task. Even a small mistake in securing data can be exploited by phishing attacks to release private information such as passwords or financial information to a malicious actor. Phishing has now proven so successful; it is the number one attack vector. Many approaches have been proposed to protect against this type of cyber-attack, from additional staff training, enriched spam filters to large collaborative databases of known threats such as PhishTank and OpenPhish. However, they mostly rely upon a user falling victim to an attack and manually adding this new threat to the shared pool, which presents a constant disadvantage in the fight back against phishing. In this paper, we propose a novel approach to protect against phishing attacks using binary visualisation and machine learning. Unlike previous work in this field, our approach uses an automated detection process and requires no further user interaction, which allows faster and more accurate detection process. The experiment results show that our approach has high detection rate.

## I. INTRODUCTION

The Internet has become an integral part of our daily activities; from communicating through social networking sites and emails to banking, studying and shopping, the Internet has touched every aspect of our life [1]. According to [2], easier access to computers, higher availability of 3G and 4G networks and the increased use of smartphones has given people the opportunity to use the internet more frequently and with more convenience. However, the increasing use of the internet has created varied opportunities to spread social engineered attacks that are designed to compromise personal information for criminal purposes [3]. Phishing has been purported as one of the greatest attack vectors that is causing great harm to online services and data security [3, 4]. This cyber-security threat attempts to trick internet users into revealing their private information such as passwords or financial account credentials, usually for the purpose of theft [3, 5].

Social engineering is the core of all phishing attacks, whether targeted or random [5]. This mechanism leads the victim to perform certain actions, such as submitting personal data directly to a malicious actor or executing malicious software that indirectly submits the data to the malicious actor without the victim knowing [4, 5]. Social engineering can manifest in the form of an email (spoofed email) or a clone of a legitimate website, so that the victim will not be able to differentiate between phishing and legitimate webpages. In addition, the attacker can use key phrases to emphasize the sense of urgency for the victim, for example," You MUST complete this account check NOW". This fear tactic leads the victim to click on a malicious link or fill out a form on the phishing site. According to the PhishLabs report [6], 255,065 unique phishing attacks were found worldwide in the third quarter of 2018, where 83.9 % of attacks targeted credentials for financial, email, cloud, payment, and SaaS services. PhishLabs affirm that this global impact of phishing attacks will continue to increase and therefore requires more efficient anti-phishing techniques to handle new and emerging phishing patterns [6]. In recent years, great effort has been directed to curb the effectiveness of phishing. A variety of approaches has been proposed including additional staff training, enriched spam filters, and large collaborative databases of known threats such as PhishTank and OpenPhish. Whilst these methods have proved effective in raising the awareness of this common cyber threat, they only have the ability to handle known phishing patterns, when a user falls victim to an attack, they manually added this new threat to the shared pool and thus leave internet users prone to new phishing attacks. However, phishers are not static in their activities; they change their mode of operation frequently to stay undetected and bypass existing techniques. When paired with the fact that cyber security is known as a reactive industry, this presents a constant disadvantage in the fight back against changing phishing patterns.

In this paper, we aim to address the aforementioned limitations by proposing a novel approach against phishing attacks using binary visualisation and machine learning. Combining the threat of phishing with machine learning and image recognition allows users to input previously unknown suspect websites and gain a better understanding of the threat in a faster time. The main contribution of this paper is an automated detection of phishing websites. As the system does not rely on multiple user verification, a faster blacklisting process can be achieved, which reduces the time for potential victims to access the link.

The remainder of this paper is organized as follows. Section 2 gives an overview of existing phishing detection techniques, their advantages and drawbacks. In Section 3, we present the methodology of the proposed method using binary visualisation and machine learning. Section 4 presents experiment results and analysis. Finally, Section 5 concludes the paper and presents future work.

## II. RELATED WORK

Over time a wide variety of approaches have been proposed to counter the ever-persistent threat of phishing in both commercial and public domains. These approaches can be classified into two main categories: user training approaches and software classification approaches [7]. Training approaches aim at raising the ability of end-users to identify phishing attacks [7, 8], which could reduce their susceptibility to falling victim to phishing attacks [8]. While classification approaches are typically designed to classify phishing and legitimate web pages on behalf of the user in an attempt to tackle issues of the human error and ignorance [7].

In this section, we focus primarily on the anti-phishing approaches that contributed to the field of phishing attack detection. Under this context, many studies based on Blacklists and Whitelists have been proposed such as Google Safe Browsing API [9], PhishNet [10], DNS-Based Blacklists and Whitelists [11], and Automated Individual White-List (AIWL) [12]. Blacklists are frequently updated lists of previously detected phishing attacks. Whitelists are lists of addresses that are considered" safe". These approaches generally have lower false positive rates [13], however, they are not efficient to protect against new phishing attacks as non-blacklisted phishing sites are not recognized [7, 13]. A study in [13] found that blacklists are only able to detect 20% of zero-hour phishing attacks, where 63% of them were blacklisted after 12 hours. However, phishing attacks are mainly performed over short periods of time. In order to avoid the blacklist draw- backs, various heuristics-based solutions have been proposed [14]. Heuristics-based approaches use the different types of characteristics that can be found in phishing attacks to define heuristics tests.

Netcraft [15] is an example of anti-phishing systems that uses heuristic methods to detect phishing attacks, with 95% accuracy. However, it is very time-consuming, even for a small dataset. SpoofGuard [16] examines phishing signatures via a list of heuristics including seen domains, URL obfuscation, nonstandard port numbers, etc. Heuristics found in the HTML content are weighted against a defined threshold value. If the weighted sum of the heuristics exceeds the defined threshold, an alarm is raised. This approach has achieved a spoof detection accuracy rate of 93.5% [17]. EarthLink toolbar [18] is a hybrid solution based on a blacklist as well as some heuristics such as the domain registration information, with 90.5% overall accuracy [17]. These solutions are more effective than blacklisting to differentiate legitimate from phishing sites. However, they are not 100% accurate since they produce low false negatives. In fact, heuristics are not guaranteed to always exist in phishing attacks, which increases the risk of misclassifying legitimate emails or websites.

Visual similarity techniques are also very useful for effectively detecting phishing websites since the phishing website is very similar to the corresponding legitimate website. These approaches compare the phishing website with the corresponding legitimate website, by using different features such as text format, HTML tags, Cascading Style Sheets (CSS) and images. If the resemblance is greater than the predefined threshold value, then it is declared phishing [19]. Phishzoo [20] was the first anti-phishing solution that was based on a visual similarity technique paired with a whitelist. This technique uses a database as a whitelist to store profiles of trusted websites. The currently visited website is analysed by matching their URL, SSL certificates, and webpage contents against stored profiles, which is an advantage over blacklist-based approaches. If the SSL certificates or addresses do not MATCH, then PhishZoo will identify the loaded website as" phishing". The main drawback of Phishzoo is the presumption that most phishing websites are simply copies of real websites. Thus, if a phishing website does not look like it is imitated (by changing the size for example), PhishZoo will prompt the user to build a new profile for that website in the whitelist.

Machine learning-based solutions try to analyse the available information of websites or webpages, by extracting static or dynamic features, and training a prediction model on a set of training data of both phishing and legitimate web pages. There is a rich family of machine learning algorithms in literature, which can be used for solving phishing detection. Authors in [21] have proposed a Neuro-Fuzzy approach to detect phishing websites, which is a combination of Fuzzy Logic and neural network. The proposed approach uses five tables of features as input of the neuro-fuzzy system including legitimate site rules, user-behaviour profile, URL information from Phishtank, etc. Then," if then rules" are generated via the neuro-fuzzy system to detect phishing. According to the tests, this approach has achieved 98.5% overall average accuracy. CANTINA+ [22] is another machine-learning framework for detecting phishing websites. This framework exploits the HTML Document Object Model (DOM), search engines and third-party services with machine learning for classification of normal and phishing sites. The classification uses a set of 15 features including IP address, page rank, embedded domain, number of dots in URL, etc. In this framework, two filters were used to increase the rate of true positives and decrease the rate of false positives. CANTINA+ achieved over 92% TP on unique phishing URLs and over 99% true positive on near- duplicate phishing URLs, and about 1.4% false positive with 20% training phish with a two-week sliding window. However, this approach suffers from performance issues due to the delay in querying from search engine.

PILFER [23] is another machine-learning approach that is proposed to detect phishing emails. In this approach, the tenfold cross-validation and random forest techniques were used for classification, support vector machines (SVM) were used for training and testing the dataset, and ten features were used to represent emails including IP-based URLs, non-matching URLs, HTML emails, number of domains, number of dots, etc. PILFER gives about 95% accuracy, but the false positive and false negative rates show that a substantial number of emails are not well classified. Moreover, many of the features cannot be extracted from old emails as the phishing sites are short-lived. In recent work [30], the author proposed a machine learning approach to detect phishing from URLs. the approach was implemented by using seven different machine learning algorithms, as Decision Tree, Adaboost, K-star, kNN(n=3), Random Forest, SMO and Naive Bayes, and different number/types of features as Natural Language Processing (NLP) based features, word vectors, and hybrid features. According to experimental results, it is concluded that the NLP based features have better performance than word vectors with an average rate of 10.86%. Additionally, the use of NLP based features and word vectors together also increases the performance of the phishing detection system with the rates of 2.24% and 13.14% according to NLP based features and the word vectors respectively.

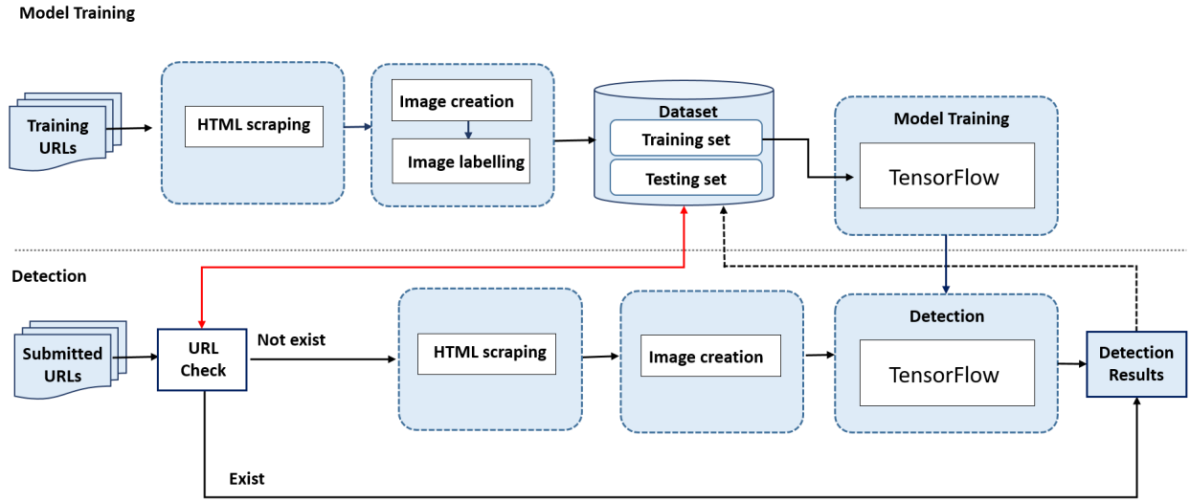

Fig. 1. Overview of the proposed approach.

In summary, most of the proposed solutions have two main issues; the first is the need for a fast access time for real-time environments and the second is the need for a high detection rate. Black-list-based solutions have fast access time, but they suffer from low detection rates, while heuristics-based and machine-learning based solutions have high detection rates but suffer from the low access time. In this paper, we aim to tackle these two main issues by combining the phishing threat with binary visualisation and machine learning. This combination can lead to faster access time with high accuracy as shown in [24]. In [24], binary visualisation and machine learning were used for malware classification with promising results. To our best knowledge, our work is the first to provide a scheme based on binary visualisation and machine learning for phishing detection. The details of the proposed system are provided in the following sections.'

## III. APPROACH OVERVIEW

In this section, we will discuss our proposed phishing detection system. As shown in Figure 1, the proposed system consists of two stages, the learning stage, and the detection stage. In the first stage, the samples and the topological structure of the machine learning TensorFlow is built, while in the second stage the submitted URLs are tested against the samples in the database to perform classification. Our approach relies on visualizing scraped HTML files onto 2D images, which are then processed by the TensorFlow that analyses them against its training modules, to distinguish between legitimate and phishing websites.

URLs passed through the system are recorded in a database, thus, each URL submitted by the user is tested to check or duplicates (*see* Figure 1). This helps in increasing the system overall performance as it could avoid the binary image reproduction process, which is a time-consuming process. If the submitted URL does not exist in the database, the system would automatically scrape the HTML code from the corresponding websites and store it in a string format. The automation of scraping the web page protects users from having to visit the potential phishing page and removes the risk of droppers and browser exploits. In addition, it prevents the user from viewing potentially inappropriate content that may be found on unknown or hacked websites. Once the website source code is scraped and stored, the corresponding binary file is passed to the image creation module, where the image visualisation method Binvis is used to convert the binary files into 2D images. Then, created images are analysed using the neural network TensorFlow to perform classification.

### A. Binary visualisation

As aforementioned, once the system scrapes and stores a target website 's source code, it then transfers it into BinVis for the RGB image creation process. Binvis is a binary data visualization tool that converts the contents of a binary file to another domain that can be visually represented (typically a two-dimensional space) [25]. This tool takes individual characters from the created string in the previous step, translates them to a binary state and then converts them to RGB values (*see* Figure 2). Binvis divided the different ASCII characters into the following classes of colours;

- Printable ASCII characters are assigned a blue colour.
- Control characters are assigned a green colour.
- Extended ASCII characters are assigned a red colour.
- Null and (non-breaking) spaces are respectively represented by black (0x00) and white (0xFF) colours.

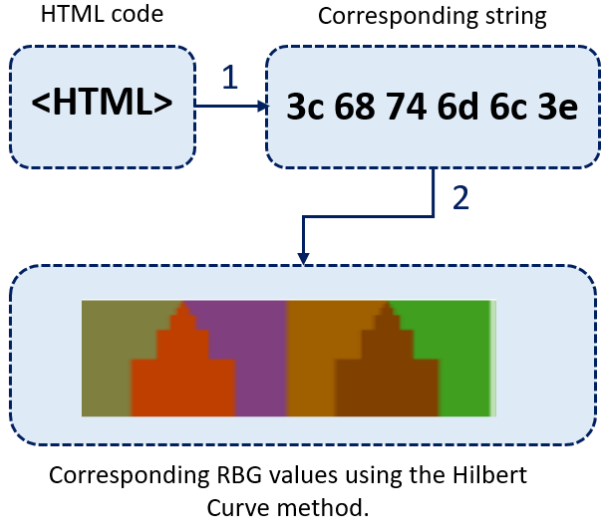

Fig. 2. Image Generation Process.

The final output of Binvis is an image with a pre-set size of 128 pixels. Figure 3 and Figure 4 show respectively Binvis images for a legitimate and a Phishing PayPal Login web page. The images were created using The Hilbert space-filling curve clustering algorithm [26], which overcomes other curves in preserving the locality between objects in multi-dimensional spaces [24, 26]. This helps to create much more appropriate RGB images for the machine learning classification process. Positive results can be concluded from Figure 3 and Figure 4 as differences between a legitimate site and its phishing counterpart were clear and apparent. The legitimate site has a more detailed RGB value because it would be constructed from additional characters sourced from licenses, hyperlinks, and detailed data entry forms. Whereas the phishing counterpart would generally contain a single or no CSS reference, multiple images rather than forms and a single login form with no security scripts. This would create a smaller data input string when scraped.

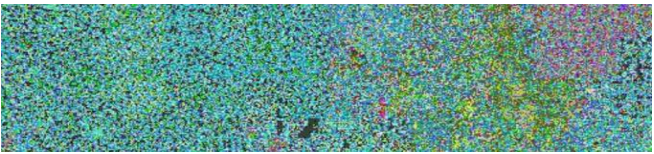

Fig. 3. Legitimate PayPal Login page.

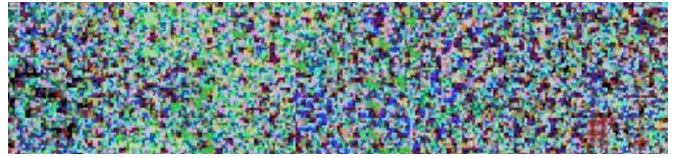

Fig. 4. Phishing PayPal Login page.

### B. Image Recognition Classifier

In order to detect phishing pages, the machine-learning algorithm TensorFlow is used to analyse and classify the Binvis images against its in-depth training. TensorFlow is flexible and it has been used for deploying machine learning systems into production across several areas of computer science, including image recognition, computer vision, robotics, information retrieval, natural language processing and geographic information extraction [27]. Its excellent image recognition ability makes it more appropriate for this application than other similar models. In fact, it could easily detect differences between the images, including differences that the human eye could not detect. TensorFlow takes as input the images produced in the previous steps to perform the classification.

For speed of testing, the convolutional neural network MobileNet [28] is employed for the retraining element. This can greatly minimize the time and space for phishing websites classification.

## IV. IMPLEMENTATION AND RESULTS

In this section, we will discuss implementation details and results of the experiments carried out over our approach in order to demonstrate its effectiveness and reliability. Especially, accuracy metric (A) was used to analyse the results and evaluate the overall performance of our approach.

$$A = \frac{TP+TN}{TP+FP+TN+FN} \quad (1)$$

Where TP is the number of instances correctly classified as phishing, TN is the number of instances correctly classified as legitimate web pages, FP is the number of instances incorrectly classified as phishing, and FN is the number of instances incorrectly classified as legitimate web pages.

Precision (P), recall (R) and f1 value (F1) metrics were also used to evaluate the performance of the classifier, where.

$$P = \frac{TP}{TP+FP} \quad (2)$$

$$R = \frac{TP}{TP+FN} \quad (3)$$

$$F1 = \frac{2 \times P \times R}{P+R} \quad (2)$$

## A. Experiment Setup

For the initial web scraping, a python script was created by using the urllib library [29]. The script would scrape the target site source code and store it in a string format. The experiments are based on the MobileNet model on python with TensorFlow open-source library. The TensorFlow framework is deployed in a virtual machine, running on Intel Core i5 CPU, 3.80 GHz, with 8 GB memory and the Ubuntu 14.04 64 bites OS. An NVIDIA GTX 1060 GPU with 6 GB memory is used as accelerator. In the training stage, the TensorFlow algorithm was trained by 250 images per category (Legitimate and phishing web pages) with a size of 128 pixels, for 4000 training steps. The learning rate was 0.005. Every image is used multiple times through training process. As shown in Table I, the phishing websites dataset contained a mixture of 25 samples from the Bank of America PHISH, PayPal Phish, ABSA Phish, DHL TRACKING Phish and Microsoft Login Phish.

TABLE I
PHISHING WEBSITES SAMPLES

| Category | Number of samples |
| --- | --- |
| Bank Of America PHISH | 5 |
| PayPal Phish | 5 |
| ABSA Phish | 5 |
| DHL TRACKING Phish | 5 |
| Microsoft Login Phish | 5 |

## A. Experimental results analysis

Several tests were carried out to determine the accuracy of the proposed classifier after the addition of more samples; five tests per trained target site were carried out to evaluate the success of the detection method. Figure 5 shows the results of the final test with the most training samples that were collected being used. It is apparent from the results that the classifier has achieved high accuracy for almost all categories, in particular, the ABSA and DHL URLs, where all submitted URLs were correctly labelled as expected. The classifier achieved lower accuracy with the PayPal URLs (85.71%), however, the precision was very high (100%).

Figure 6 shows the overall results of the proposed approach, which achieved an overall detection accuracy of 94.16%, which is high and meets the required accuracy rate in practical use. The precision of the classification is also very high with a rate of 95.83%, which shows strong overall confidence in the pattern recognition process. This accuracy rate is interpreted as an acceptable and good result for phishing detection. The recall rate was lower than the precision rate (87.50%) because of the PayPal results that need further investigation in future works.

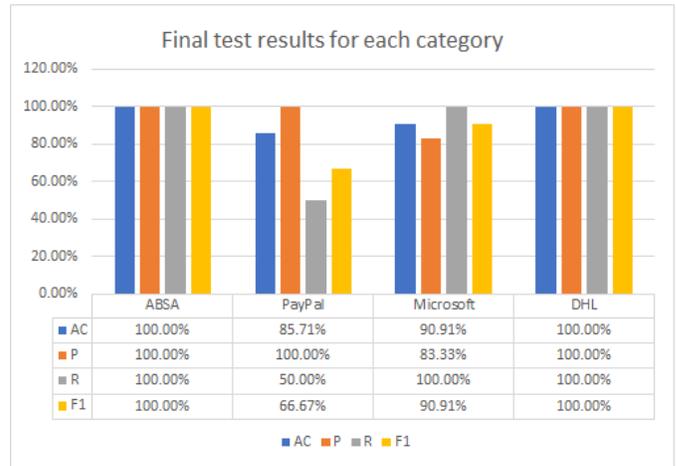

Fig. 5. Final test results by category.

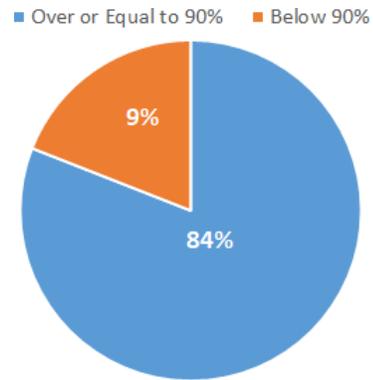

Fig. 6. Overall results for the final test

## V. CONCLUSION

Phishing has become a serious threat in online space, largely driven by evolving web, mobile, and social networking technologies. Due to the rapid spreading of new phishing websites and distributed phishing attacks, current phishing detection techniques need to be greatly enhanced to effectively combat emerging phishing attacks. In this paper, we have proposed a novel phishing detection method, leveraging multilevel artificial intelligence that uses a combination of neural network paired with a binary visualization. Using visual representation techniques allows to obtain an insight into the structural differences between legitimate and phishing web pages. From our initial experimental results, the method seems promising and being able to fast detection of phishing attacker with high accuracy. Moreover, the method learns from the misclassifications and improves its efficiency.

In the future, we plan to improve this work by the use of more samples for training and testing and utilising GPU for binary visualization and CNN classification, which will with no doubt enhance the predictive accuracy of the classifier. Furthermore, we intend to apply the proposed solution with more languages such as Russian, Greek and Chinese languages, and trained the system on 404-error HTML code.

ACKNOWLEDGMENT

This project has received funding from the European Unions Horizon 2020 research and innovation programme under grant agreement no. 786698. This work reflects authors view and Agency is not responsible for any use that may be made of the information it contains.